\documentclass[pdflatex,sn-mathphys-num]{sn-jnl}

\usepackage{url}
\usepackage{hyperref}
\usepackage{microtype}
\usepackage{xspace}
\usepackage{balance}
\usepackage{amsmath}
\usepackage{amssymb}
\usepackage{siunitx}
\usepackage{pifont}
\usepackage{booktabs}
\usepackage{multirow}
\usepackage{tabularx}
\usepackage{array}
\usepackage{colortbl}
\usepackage{graphicx}
\usepackage{subcaption}
\usepackage{float}
\usepackage{pdflscape}
\usepackage[table]{xcolor}
\usepackage{enumitem}
\usepackage{acronym}
\usepackage[strict]{changepage}
\usepackage{framed}
\usepackage{tcolorbox}
\usepackage{adjustbox}
\definecolor{checkgreen}{RGB}{34,139,34}
\definecolor{crossred}{RGB}{160,160,160}
\definecolor{orange}{RGB}{255,165,0}  

\definecolor{promptbg}{HTML}{B7E0FF}
\definecolor{promptborder}{HTML}{024CAA}

\definecolor{answerbg}{rgb}{0.85,1,0.85}
\definecolor{answerborder}{rgb}{0.0,0.6,0.30}

\newcommand{\wcircle}[1]{\ding{\numexpr171 + #1}}
\newcommand{\bcircle}[1]{\ding{\numexpr181 + #1}}
\newcommand{\cmark}{\textcolor{checkgreen}{\ding{51}}}
\newcommand{\xmark}{\textcolor{crossred}{\ding{55}}}

\newenvironment{prompt}{%
  \MakeFramed{\advance\hsize-\width\FrameRestore}%
  \noindent\hspace{-4.55pt}%
  \begin{adjustwidth}{}{7pt}%
}{%
  \end{adjustwidth}%
  \endMakeFramed%
}

\newenvironment{answer}{%
  \MakeFramed{\advance\hsize-\width\FrameRestore}%
  \noindent\hspace{-4.55pt}%
  \begin{adjustwidth}{}{7pt}%
}{%
  \end{adjustwidth}%
  \endMakeFramed%
}

\usepackage{ifthen}
\newboolean{showcomments}

\setboolean{showcomments}{true}

\ifthenelse{\boolean{showcomments}}
{%
  \newcommand{\mynote}[2]{%
    \fbox{\bfseries\sffamily\scriptsize #1}
    {\small$\blacktriangleright$\textsf{\emph{#2}}$\blacktriangleleft$}
  }
}
{%
  \newcommand{\mynote}[2]{}
}

\begin{document}
\title[Evaluating LLMs for Obfuscation Detection and Classification in Android Apps]{Evaluating LLMs for Obfuscation Detection and Classification in Android Apps}

\author*[1]{\fnm{Luca} \sur{Ferrari}}\email{luca.ferrari@imtlucca.it}
\equalcont{These authors contributed equally to this work.}

\author[2]{\fnm{Marco} \sur{Alecci}}\email{marco.alecci@uni.lu}
\equalcont{These authors contributed equally to this work.}

\author[2]{\fnm{Jordan } \sur{Samhi}}\email{jordan.samhi@uni.lu}
\author[2]{\fnm{Tegawendé F.}\sur{Bissyandé}}\email{tegawendé.bissyandé@uni.lu}
\author[2]{\fnm{Jacques } \sur{Klein}}\email{jacques.klein@uni.lu}
\author[3]{\fnm{Mariano } \sur{Ceccato}}\email{mariano.ceccato@univr.it}
\author[4]{\fnm{Luca } \sur{Verderame}}\email{luca.verderame@unige.it}

\affil*[1]{\orgdiv{IMT School for Advanced Studies Lucca }, \orgname{Organization}, \orgaddress{\street{Piazza S.Francesco, 19}, \city{Lucca LU}, \postcode{55100}, \state{Italy}, \country{Italy}}}

\affil[2]{\orgdiv{Interdisciplinary Centre for Security, Reliability and Trust}, \orgname{Organization}, \orgaddress{\street{29 Av. John F. Kennedy}, \city{Luxembourg}, \postcode{1855}, \state{Luxembourg}, \country{Luxembourg}}}

\affil[3]{\orgdiv{University of Verona}, \orgname{Organization}, \orgaddress{\street{Str. le Grazie 15}, \city{Verona}, \postcode{37134}, \state{Italy}, \country{Italy}}}
\affil[4]{\orgdiv{University of Genoa}, \orgname{Organization}, \orgaddress{\street{V. all'Opera Pia 13}, \city{Genova}, \postcode{16145}, \state{Italy}, \country{Italy}}}

\maketitle
\begin{abstract}

Android applications (apps) developers increasingly rely on code obfuscation techniques to hinder reverse engineering and protect intellectual property. However, obfuscation also reduces the effectiveness of static analysis and vulnerability detection tools, creating challenges for Android security analysis. Existing approaches for detecting obfuscation in Android apps predominantly rely on handcrafted heuristics, engineered features, or task-specific learning pipelines, which may struggle to generalize across evolving obfuscation strategies.

This paper presents a large-scale empirical study investigating the capability of Large Language Models (LLMs) to detect obfuscation in Android apps through semantic reasoning. 
Our study evaluates whether off-the-shelf LLMs can identify obfuscated code without relying on handcrafted rules, predefined signatures, or dedicated model training. The empirical evaluation is conducted on both a controlled benchmark containing an app obfuscated with multiple techniques and a real-world dataset of Android apps collected from Google Play. The study further examines the impact of prompt design, model selection, and decision thresholds across several open-weight and proprietary LLMs. Finally, the analysis compares LLM-based reasoning with existing SAST-based obfuscation-detection approaches and discusses the broader implications and limitations of applying LLMs to Android security analysis.
\end{abstract}

\section{Introduction}
\label{sec:introduction}


Android applications (apps) constitute one of the most widespread software ecosystems, with billions of users relying on mobile apps distributed through marketplaces such as Google Play. However, the inherent openness of this ecosystem allows any actor to download, reverse engineer, and inspect application code, enabling large-scale vulnerability discovery, intellectual property theft, and malicious repackaging~\cite{zhou2012dissecting, li2017androidrepackaging, wu2026android}. This threat model has driven developers to adopt \textit{code obfuscation} as a primary line of defense~\cite{kargen2023characterizing}.

Code obfuscation comprises a set of semantics that preserve transformations that deliberately degrade code readability and analyzability~\cite{collberg1997taxonomy}. Techniques such as identifier renaming and control-flow modification are widely used to hinder both human reverse engineering and automated analysis~\cite{conti2022obfuscation}. Recent large-scale empirical studies reveal that a substantial fraction of Android apps are obfuscated, with a steadily increasing adoption trend across years and app categories~\cite{GooglePlayObfuscation}. This widespread use has effectively reshaped the landscape of Android security analysis. Despite its defensive intent, obfuscation introduces a fundamental tension: it simultaneously protects apps and undermines the effectiveness of security analysis techniques. Prior work has shown that obfuscation can significantly reduce the detection capability of state-of-the-art Static Application Security Testing (SAST) tools, which rely on structural patterns, control-flow graphs, and syntactic heuristics to identify vulnerabilities~\cite{conti2022obfuscation}. As a result, both defenders and analysts face increasing difficulty in assessing the true security posture of Android apps.

The literature has extensively studied \wcircle{1} the prevalence and evolution of obfuscation in the wild~\cite{GooglePlayObfuscation}, and \wcircle{2} its impact on vulnerability detection pipelines~\cite{conti2022obfuscation}. In contrast, the problem of automatically detecting whether an APK is obfuscated, robustly and at scale, remains underexplored. Current approaches primarily rely on handcrafted features or rule-based heuristics~\cite{mirzaei2018androdet, wang2017obfuscator}, which are inherently brittle and struggle to generalize across diverse and evolving obfuscation strategies. Recent advances in \textit{Large language models (LLMs)} have demonstrated strong capabilities across a wide range of code-related tasks, including bug detection and fixing~\cite{feng2024prompting, li2024enhancing, bouzenia2024repairagent, kang2023large, jiang2026survey}, testing~\cite{liu2024make, huang2024crashtranslator, chen2024chatunitest, schafer2023empirical}, vulnerability detection~\cite{zhao2025apppoet, qian2025lamd, sun2024gptscan, guo2024outside, sun2025raml}, and other applications~\cite{khare2023understanding, pei2023can, ma2023lms, sun2023automatic, zhong2024can, alecci2025toward, alecci2026replacing}. These properties make them particularly well-suited to capture obfuscation patterns that transcend syntactic transformations and evade traditional static analysis techniques. However, the extent to which LLMs can semantically recognize and distinguish obfuscation techniques in Android apps remains unexplored.

In this paper, we conduct the first large-scale empirical study of the ability of off-the-shelf LLMs to reason about code obfuscation in Android apps. In particular, we investigate whether LLMs can \wcircle{1} detect whether an APK contains obfuscated code, and \wcircle{2} identify the specific obfuscation techniques being applied. To support this study, we design an LLM-driven analysis pipeline that operates directly on Smali code extracted from Android APKs, enabling semantic inspection of obfuscation patterns without relying on handcrafted rules, predefined feature sets, dedicated pre-training, or access to the app's original source code. Unlike prior feature-engineered or trained approaches, our study evaluates whether modern off-the-shelf LLMs can generalize across heterogeneous obfuscation transformations purely through semantic reasoning.

To investigate these questions, we conducted an empirical evaluation involving multiple state-of-the-art open-weight and proprietary LLMs, three prompt-engineering strategies, and ten Android obfuscation techniques. Our results show that LLMs can effectively recognize obfuscation patterns directly from Smali code, achieving strong detection and classification performance while substantially outperforming existing SAST-based approaches. While identifier-renaming transformations were reliably recognized, techniques such as reflection and call indirection remained more difficult to detect. We further evaluate the most effective configurations on a large-scale dataset of real-world Android applications collected from Google Play, where manual validation indicates an F1-score of 0.88, demonstrating the applicability of our findings beyond controlled experimental settings.

\noindent
\textbf{Contributions.}
This paper makes the following contributions:
\begin{itemize}[leftmargin=*]
    \item We present the first empirical study investigating the capability of off-the-shelf LLMs to detect and classify code obfuscation techniques in Android apps.
    \item We provide a comprehensive review of the Android obfuscation landscape, including existing obfuscation-detection approaches, commonly adopted obfuscation techniques, and the tools most widely used to obfuscate Android apps. 
    \item We conduct an extensive empirical evaluation of multiple LLMs, prompt formulations, and obfuscation techniques on both a controlled benchmark with known ground truth and a large-scale dataset of real-world Android apps collected from Google Play.
    \item We compare LLMs against traditional rule-based and SAST-based approaches for obfuscation detection. 
    \item We publicly release our benchmark,  prompts, implementation artifacts, and experimental framework to support reproducibility and future research on Android obfuscation analysis.
\end{itemize}

\noindent
\textbf{Data Availability.} We publicly release all associated resources to facilitate further research: 
\begin{center}
   \url{https://github.com/Mobile-IoT-Security-Lab/LLMObfuscDetection}
\end{center}
\section{Background: Android Obfuscation Landscape}
\label{sec:background}
In this section, we review the Android obfuscation landscape, including the structure of Android apps, commonly adopted obfuscation techniques, and widely used obfuscation tools. This background provides the foundation for our empirical study of LLM-based obfuscation analysis.

\subsection{The Android Ecosystem.}
Modern Android apps are primarily developed in Java and Kotlin and are built around four core components: activities (UI), services (background tasks), broadcast receivers (event handling), and content providers (data sharing). These are packaged into an Android Package (APK), a ZIP archive containing code and resources. An APK includes key elements such as the \texttt{AndroidManifest.xml} (metadata and permissions), \texttt{classes.dex} files (Dalvik bytecode), resource directories (\texttt{assets}, \texttt{res}), native libraries (\texttt{lib}), and \texttt{resources.arsc}, which links code to resources. The \texttt{META-INF} directory stores signatures for integrity verification. Obfuscation is typically applied before generating \texttt{classes.dex}, and may also affect other components.

APKs are distributed via app stores like Google Play~\cite{GooglePlay}, making them easily accessible for analysis. Tools such as Ghidra~\cite{ghidra}, Apktool~\cite{apktool}, and jadx~\cite{jadx} enable reconstruction of source-like code from bytecode, raising concerns about code exposure. To mitigate this, developers employ obfuscation techniques to protect proprietary logic.

\subsection{Code Obfuscation Techniques}
Obfuscation refers to a broad class of program transformations aimed at reducing code comprehensibility for human analysts while preserving the original behavior of the app. Over the years, numerous studies have categorized and examined these techniques in depth \cite{dong2018understanding,wermke2018large,zhang2021android,conti2022obfuscation,guo2022survey}. In the context of Android applications, several obfuscation strategies are particularly widespread, following the taxonomy introduced by Niroshan et al.~\cite{niroshan2025empirical}:

\noindent
\textbf{\bcircle{1} Identifier Renaming (IR)}
One of the simplest and most commonly adopted techniques consists of substituting meaningful identifiers, such as class names, method names, and variables, with arbitrary or non-descriptive strings. This transformation removes semantic cues from the code, making manual inspection significantly more difficult while leaving execution unaffected.

\noindent
\textbf{\bcircle{2} Control Flow Transformation (CF)}
Control flow obfuscation aims to alter the apparent structure of program execution in order to complicate both static analysis and reverse engineering \cite{zhang2021android,guo2022survey}. This is typically achieved through multiple mechanisms:
\begin{itemize}[leftmargin=*]
\item \textit{Control Flow Flattening}: The original structure of a function is reorganized into a single dispatcher loop (often implemented with a \texttt{switch} construct), where execution order is governed by a control variable. Although semantically equivalent to the original code, the resulting structure appears highly non-linear and difficult to interpret.
\item \textit{Call Indirection}: Direct method invocations are replaced with intermediate wrapper functions that forward the execution to the actual target. This additional level of indirection obscures relationships between components and complicates the reconstruction of the call graph \cite{zhang2021android,bacci2018detection}.
\item \textit{Reflection}: By relying on the \texttt{java.lang.reflect.*} API, applications can defer method resolution to runtime. Since the actual targets of these calls are determined dynamically, static analysis tools often fail to accurately model the program behavior.
\item \textit{Junk Code and Opaque Predicates}: Another common approach involves inserting irrelevant instructions (e.g., \texttt{nop}) or misleading control structures. In particular, opaque predicates, conditions whose outcome is predetermined but non-trivial to infer statically, are used to introduce branches where only one path is feasible, while the other leads to non-executable or meaningless code \cite{guo2022survey,li2019obfusifier}.
\end{itemize}

\noindent
\textbf{\bcircle{3} String Encryption (SE)}
Sensitive information embedded in applications, such as API keys or descriptive strings, can be exploited during reverse engineering if stored in plaintext. To counter this, obfuscation techniques encode such strings into encrypted representations, which are then decoded dynamically at runtime. This approach prevents straightforward extraction through static inspection.

\subsection{Code Obfuscation Tools}
Code obfuscation tools are essential in mobile app security to protect applications from reverse engineering by renaming classes, methods, and variables into meaningless identifiers, optimizing code size, and adding layers of complexity that hinder static and dynamic analysis. These tools are particularly relevant in Android app ecosystems, where APK analysis is common in security research. Over the years, both commercial and non-commercial solutions have been developed to address these needs, as shown in Table~\ref{tab:code-obfuscation-tools}.

\begin{table}[ht]
\centering
\caption{Code Obfuscation Tools}
\label{tab:code-obfuscation-tools}
\small
\setlength{\tabcolsep}{6pt}
\renewcommand{\arraystretch}{1.12}
\begin{tabularx}{\linewidth}{@{}l >{\raggedright\arraybackslash}X c@{}}
    \toprule
    \textbf{Tool} & \textbf{Methodology} & \textbf{Open Source} \\
    \midrule
    ProGuard/R8~\cite{proguard,R8} & Renaming, shrinking, optimization & \cmark \\
    Allatori~\cite{allatori} & String encryption, control flow, anti-debug & \xmark \\
    DashO~\cite{DashO} & R8 integration, string encryption & \xmark \\
    Obfuscapk~\cite{obfuscapk} & Multi-type obfuscation & \cmark \\
    DexGuard~\cite{dexguard} & RASP, polymorphic, cryptography & \xmark \\
    \bottomrule
\end{tabularx}
\vspace{0.4em}
\end{table}

In the following, we briefly describe the most representative obfuscation tools:
\begin{itemize}[leftmargin=*]
\item \textbf{ProGuard/R8} is an open-source tool widely used for Java and Android bytecode shrinking, optimization, and obfuscation. R8 is the successor to ProGuard and has been the default Android code shrinker and obfuscator since its integration into the Android build pipeline, while remaining compatible with existing ProGuard rules. It replaces meaningful names with short, meaningless ones while preserving functionality, making decompiled code harder to understand.
\item \textbf{Allatori} is a commercial Java obfuscator offering advanced features like control flow obfuscation, string encryption, and anti-debugging techniques. It is optimized for desktop Java applications but applicable to Android as well.
\item \textbf{DashO}, from PreEmptive Solutions, is a commercial obfuscator for Java and Android that supports string encryption, code virtualization, and integration with R8/ProGuard for enhanced protection.
\item \textbf{Obfuscapk} is a specialized tool for obfuscating Android APK files, focusing on targeted transformations for package-level protection. 
\item \textbf{DexGuard}, developed by Guardsquare, is a commercial evolution of ProGuard tailored for Android, featuring runtime application self-protection (RASP), polymorphic obfuscation, data encryption, and native code hardening.
\end{itemize}
\section{Related Work}
\label{sec:relatedwork}

Several approaches have been proposed in the literature to detect and analyze code obfuscation in Android apps, spanning static and dynamic analysis techniques, as well as traditional machine learning and, more recently, deep learning-based methods. In parallel, large-scale empirical studies have highlighted the widespread adoption of obfuscation in real-world applications, underscoring the need for scalable, generalizable detection techniques. 
These approaches can be categorized into three main categories based on their primary objective: \wcircle{1} \textit{Detection methods}, which focus solely on identifying the presence of obfuscation regardless of the app's intent; \wcircle{2} \textit{Malware-oriented approaches}, which target obfuscation specifically in the context of malicious applications to evade anti-malware detection; and \wcircle{3} \textit{Static Analysis techniques}, which emphasize analyzing obfuscated code structure and semantics without execution, often as a preprocessing step for further analysis. 
In the following, we review the most relevant works in this domain, organizing them according to their methodological foundations and discussing their main limitations. Table~\ref{tab:obfuscation_tools} summarizes the main approaches proposed in the literature for Android obfuscation detection, together with their underlying methodologies, code availability, and their categorization within this taxonomy.

\begin{table}[ht]
\centering
\scriptsize
\caption{Existing approaches for Android obfuscation detection.}
\label{tab:obfuscation_tools}
\setlength{\tabcolsep}{4pt}
\renewcommand{\arraystretch}{1.12}
\begin{tabularx}{\linewidth}{@{}c >{\raggedright\arraybackslash}p{2.6cm} X c c >{\raggedright\arraybackslash}p{1.8cm}@{}}
    \toprule
    \textbf{Year} & \textbf{Tool / Authors} & \textbf{Methodology} & \textbf{Category} & \textbf{Code Available} & \textbf{Venue} \\
    \midrule
    2017 & Wang et al.~\cite{wang2017changed} & Supervised learning & Detection & \cmark & MOBILESoft \\
    2018 & Martinelli et al.~\cite{martinelli2018evaluating} & Model checking & Detection & \xmark & JPDC \\
    2018 & Obfuscan~\cite{wermke2018large} & Static analysis & Detection & \xmark & ACSAC \\
    2019 & AndrODet~\cite{mirzaei2019androdet} & Online learning & Detection & \cmark & FGCS \\
    2020 & DroidPDF~\cite{sun2020droidpdf} & Machine learning & Detection & \xmark & IEEE Access \\
    2020 & EspyDroid~\cite{gajrani2020espydroid+} & Hybrid analysis & Malware & \cmark & COSE \\
    2020 & DINA~\cite{alhanahnah2020dina} & Hybrid analysis & Malware & \cmark & TIFS \\
    2020 & StaDART~\cite{ahmad2020stadart} & Hybrid analysis & Malware & \cmark & JSS \\
    2020 & DANdroid~\cite{millar2020dandroid} & Deep learning & Detection & \xmark & CODASPY \\
    2022 & Conti et al.~\cite{conti2022obfuscation} & Deep learning & Malware & \cmark & JISA \\
    2022 & APKHUNT~\cite{apkhunt} &  Static analysis & Static analysis & \cmark & -- \\
    2022 & Trueseeing~\cite{trueseeing} &  Static analysis & Static analysis & \cmark & -- \\
    2023 & SEBASTiAn~\cite{pagano2023sebastian} &  Static analysis & Static analysis & \cmark & SoftwareX \\
    2025 & Niroshan et al.~\cite{niroshan2025empirical} & Machine learning & Detection & \cmark & SAC \\
    \bottomrule
\end{tabularx}
\end{table}

\textbf{Obfuscation Detection.}
Early approaches to obfuscation detection rely on handcrafted features combined with traditional machine learning techniques. AndrODet~\cite{mirzaei2019androdet} represents one of the most advanced systems in this category, leveraging static features extracted from bytecode alongside online learning algorithms to identify different obfuscation techniques. It achieves an accuracy of 92.02\% for identifier renaming, 81.41\% for string encryption, and 68.32\% for control-flow obfuscation. However, its effectiveness heavily depends on feature engineering and degrades when dealing with more complex transformations, particularly control-flow obfuscation. Wang et al.~\cite{wang2017changed} propose an early method focused on identifying the obfuscation tool used (e.g., ProGuard, Allatori, DashO), achieving high accuracy (0.97).
Nevertheless, the approach targets tool attribution rather than the detection or characterization of specific obfuscation techniques, making it not directly comparable with fine-grained detection methods. Other works focus on specific obfuscation strategies. Martinelli et al.~\cite{martinelli2018evaluating} employ model checking techniques to detect control-flow obfuscation patterns, but their method is limited to a single transformation class. Similarly, Sun et al.~\cite{sun2020droidpdf} introduce an entropy-based approach that captures general anomalies indicative of obfuscation, yet it does not distinguish among different obfuscation techniques. Obfuscan~\cite{wermke2018large} employs black-box static analysis on APKs to determine whether an application is obfuscated. It inspects package, class, method, and field names, method overloading, the presence or absence of debug information, source files, and annotations, using heuristics derived largely from ProGuard. 

More recent work explores deep learning-based solutions. Conti et al.~\cite{conti2022obfuscation} propose a hybrid model combining heterogeneous representations, including natural language-inspired features and image-based encodings of code. Their approach achieves high performance, with an average F-measure of 0.985 across multiple obfuscation classes, demonstrating the benefits of multi-representation learning. However, these methods still rely on carefully engineered input representations and may struggle to generalize to unseen or evolving obfuscation strategies. In parallel, large-scale empirical studies have analyzed the prevalence of obfuscation in real-world applications. Niroshan et al.~\cite{niroshan2025empirical} examine over \num{500000} Android applications from Google Play, showing that obfuscation adoption has increased significantly over time and is particularly prevalent among popular applications. These findings highlight both the diversity and the continuous evolution of obfuscation techniques, reinforcing the need for flexible and adaptive detection approaches.

\noindent
\textbf{Obfuscation in Malware Analysis.}
Several works incorporate obfuscation analysis within broader malware detection pipelines. DANdroid~\cite{millar2020dandroid}, for instance, combines multiple machine learning models to detect malware while identifying obfuscation-related artifacts such as class encryption, API call obfuscation, and string encryption. Other studies~\cite{gajrani2020espydroid+,alhanahnah2020dina,ahmad2020stadart} adopt dynamic or hybrid analysis techniques to inspect runtime behavior, particularly targeting malware that employs reflection, dynamic code loading, or class encryption.

\noindent
\textbf{Static Analysis Tools with Obfuscation Detection.}
In addition to research-oriented approaches, several static analysis tools designed for vulnerability detection provide partial support for identifying obfuscation patterns. Tools such as \textit{Sebastian}~\cite{pagano2023sebastian}, \textit{APKHunt}~\cite{apkhunt}, and \textit{Trueseeing}~\cite{trueseeing} are primarily SAST (Static Application Security Testing) frameworks that perform static code inspection to detect security vulnerabilities. \textit{Sebastian} targets large-scale analysis by combining multiple detectors to identify issues such as insecure data flows, API misuse, and privacy leaks. Within this context, it can also surface indicators associated with obfuscation, including reflection usage, dynamic code loading, and anomalous control-flow structures. However, these signals are indirect and not specifically designed for systematic obfuscation detection. \textit{APKHunt} follows a rule-based approach with an extensible plugin architecture, enabling the identification of suspicious patterns such as identifier renaming, string encryption, and packing techniques. These checks are heuristic in nature and primarily support vulnerability analysis rather than providing a structured characterization of obfuscation techniques. Similarly, \textit{Trueseeing} offers a lightweight static inspection framework for reverse engineering and security analysis. It incorporates pattern-based mechanisms to identify artifacts such as code shrinking, control-flow alterations, and encrypted resources, typically as part of broader analysis workflows. Despite their practical relevance, these tools are not designed as dedicated obfuscation detectors. Their analyses are generally heuristic-driven, coarse-grained, and not evaluated against formal obfuscation taxonomies. Consequently, they provide auxiliary signals rather than a comprehensive and fine-grained classification of obfuscation techniques.

\noindent
\textbf{Research Gap.}
To the best of our knowledge, current approaches to obfuscation detection are largely constrained to feature-based, heuristic-driven, or representation-dependent paradigms. While effective in controlled settings, these methods may struggle to generalize across diverse and evolving obfuscation strategies. In particular, the potential of LLMs for reasoning over program structure and semantics in the context of obfuscation detection remains largely unexplored. This gap motivates our work, where we investigate whether LLMs can provide a more flexible, scalable, and generalizable solution for detecting obfuscation in Android applications.
\section{Experimental Setup.}
\label{sec:experimentalSetup}
In this section, we present our experimental setup, outlining our methodology, implementation details, datasets, and metrics used in our experiments.

\subsection{Methodology}
\label{sec:methodology}
\begin{figure}[ht]
    \centering
    \includegraphics[width=\linewidth]{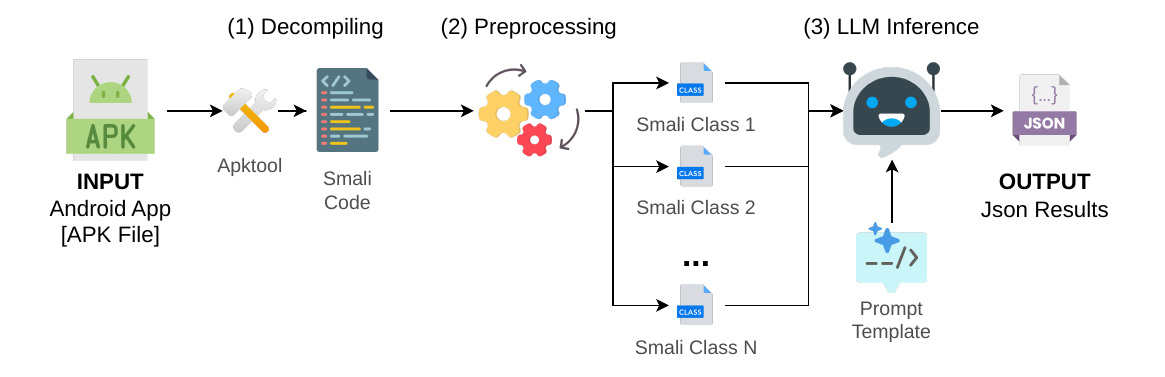}
    \caption{Overview of the LLM-based experimental pipeline for detecting obfuscation in Android apps.}
    \label{fig:approachOverview}
\end{figure}

Our evaluation aims to assess whether off-the-shelf LLMs can automatically detect and characterize obfuscation in Android apps. To this end, we design an LLM-based pipeline that operates directly on Smali code extracted from APK files. The pipeline consists of three main phases: \bcircle{1} decompilation, \bcircle{2} preprocessing, and \bcircle{3} LLM inference. An overview of the pipeline is shown in Figure~\ref{fig:approachOverview}.

\noindent
\textbf{\bcircle{1} Decompilation.}
In the first phase, the input APK is decompiled using \texttt{apktool}~\cite{apktool}. This process extracts the APK into a directory structure containing the manifest, resources, and code artifacts, while converting the compiled Dalvik bytecode stored in \texttt{.dex} files into human-readable Smali code. Since Android obfuscation techniques are primarily applied to the generated bytecode, we perform our analysis directly on Smali code to evaluate whether LLMs can identify obfuscation patterns directly from bytecode-level representations without requiring access to the original source code, which is typically unavailable in practice.


\noindent
\textbf{\bcircle{2} Preprocessing.}  
\label{sec:preprocessing}
The output produced by apktool typically contains a large number of \texttt{.smali} files with highly variable sizes and content. Since our analysis targets developer-implemented logic, we first filter out Smali code belonging to Android framework components and third-party libraries, leveraging AndroLibZoo, the Android library dataset provided by Samhi et al.~\cite{AndroLibZoo}. This choice is motivated by two considerations: First, third-party libraries are often reused across many apps and may already contain obfuscated code unrelated to the developer's protection strategy. Second, our objective is to determine whether the app itself has been obfuscated, as reflected primarily in the code developed and packaged by the app authors. Importantly, removing library code does not eliminate interactions with external components, as developer-defined classes still contain calls to framework and library APIs. 

After filtering, the remaining Smali files are reorganized at the class level, which represents the basic unit of analysis in our pipeline. Nevertheless, even after removing framework and third-party libraries, modern Android apps may still contain hundreds or even thousands of developer-defined classes. Performing LLM inference for every class would therefore introduce a substantial computational burden, significantly increasing execution time and monetary costs. More generally, prior work has shown that naively submitting large portions of an APK to LLMs does not scale well because of token limitations, inference latency, and API costs~\cite{qian2025lamd, alecci2025toward, sun2025raml}. 

To address this challenge, we adopt a statistically significant random sampling strategy rather than exhaustively analyzing all classes. Similar sampling-based methodologies are commonly employed in empirical software engineering studies when validating large datasets or performing manual inspections, where analyzing the entire population would be prohibitively expensive~\cite{li2026automatically, guglielmi2024help, alecci2025damflow, zhang2025killing, wang2024makes,  alecci2026taskflow, alecci2024improving, alecci2026geotwins}. The underlying rationale is that a sufficiently large and representative sample can provide reliable estimates of the characteristics of the whole population while substantially reducing the required effort.
Specifically, for each app, we compute the sample size using the standard finite-population sample size formula with a confidence level of 95\% and a margin of error of 5\%. The sample size is computed using Cochran's sample size formula with finite population correction~\cite{cochran1977sampling}. Consequently, the number of analyzed classes varies across apps according to their size, while remaining statistically representative of the overall set of developer-defined classes. This sampling strategy represents a practical trade-off between analysis depth and scalability: larger samples provide more complete coverage of the application at the expense of higher computational cost, whereas smaller samples reduce execution time while maintaining a quantifiable level of statistical confidence. The sampled classes are then independently processed during the LLM inference phase.

\noindent
\textbf{\bcircle{3} LLM Inference.}  
For each class in the selected sample, we perform independent LLM inference. In the binary obfuscation detection task, the LLM receives the class name, the corresponding Smali code, and a zero-shot prompt asking whether the class shows signs of obfuscation. We design three prompt variants, denoted as $prompt_{v1}$, $prompt_{v2}$, and $prompt_{v3}$, which progressively increase the amount of domain knowledge provided to the model. The first prompt provides only a minimal task description, the second introduces a formal definition of obfuscation, and the third explicitly enumerates the obfuscation techniques considered in this study.  This design allows us to systematically evaluate, in RQ1, (see Section~\ref{sec:rq1}, the impact of prompt specificity on obfuscation detection performance. The complete prompts are reported in Appendix~\ref{app:prompts} and are also available in our repository

The output of this phase is the proportion of sampled classes classified as obfuscated. This value can already be interpreted as an app-level obfuscation score, representing the extent to which the analyzed classes exhibit obfuscation patterns. In addition, when a binary app-level decision is required, we apply a decision threshold $T$ to this score. An app is classified as obfuscated if the ratio of classes flagged as obfuscated exceeds $T$:

\begin{equation}
\label{eq}
\frac{N_{\text{obfuscated}}}{N_{\text{sample}}} > T
\end{equation}

\noindent
In RQ1 (see Section~\ref{sec:rq1}), we empirically evaluate nine threshold values $[0.1-0.9]$ to identify the operating point that provides the best trade-off between detection capability and false-positive control. 

Furthermore, the LLM-driven pipeline is prompt-agnostic and can be adapted to related tasks. For instance, we also define a dedicated prompt for identifying the specific obfuscation technique applied to a class (see RQ2, Section~\ref{sec:rq2}). This classification prompt is reported in Appendix~\ref{app:prompts}. The output of this task is a ranked list of candidate obfuscation techniques, ordered according to the model's confidence.

Overall, the proposed pipeline balances scalability, cost, and analytical accuracy by combining statistically grounded sampling with LLM-based semantic reasoning, enabling the efficient analysis of large Android app datasets.

\subsection{Implementation.}
We implemented the proposed methodology using a set of Python scripts that are publicly available in our replication package.

To evaluate the effectiveness of LLMs for Android obfuscation detection, we consider six state-of-the-art models spanning both open-weight and proprietary families:
\begin{enumerate}[leftmargin=*]
\item \textit{\textbf{Local open-weight models}}, representing a privacy-preserving deployment setting where inference is performed entirely on local hardware. This category includes \textit{deepseek-r1:32b}, \textit{gemma3:27b}, \textit{gpt-oss:20b}, and \textit{qwen3:30b}.
\item \textit{\textbf{Cloud proprietary models}}, representing commercially deployed state-of-the-art systems accessed through remote APIs. This category includes \textit{gpt-4o-mini} and \textit{gpt-5-mini}.
\end{enumerate}

These models were selected because they represent some of the most widely adopted and recognizable systems in the current LLM landscape, enabling a meaningful comparison across different model families, deployment settings, and parameter scales. Later in RQ1 (see Section~\ref{sec:rq1}), we systematically compare all models (on top of prompt variants and decision thresholds) to identify the configuration that achieves the best obfuscation detection performance.

Given the known non-deterministic behavior of LLMs~\cite{tosem2025}, which is further discussed later in our Limitations Section (see Section~\ref{sec:limitations}), we adopted several measures to improve the robustness and reproducibility of our experiments. First, each prompt is executed 5 times per analyzed class, and the majority label is retained as the final class-level prediction. This procedure is intended to mitigate output variability and reduce the impact of occasional hallucinations or inconsistent responses. Second, all models were evaluated with a temperature of 0 to reduce output variability. Third, all interactions were executed in a stateless manner, with each prompt submitted independently and without retaining conversational history. Together, these measures reduce the impact of stochastic outputs and ensure that all models are evaluated under consistent experimental conditions.

\subsection{Datasets.}
\label{sec:datasets}
The experimental methodology is supported by two distinct datasets: \bcircle{1} $D_{GT}$ and \bcircle{2} $D_{W}$, which will be described hereafter:

\noindent
\textbf{\bcircle{1} The first dataset, $D_{GT}$}, consists of 10 non-obfuscated APKs with available source code, randomly selected from F-Droid~\cite{fdroid}. Each app was manually obfuscated using \wcircle{1} nine \textsc{Obfuscapk} transformations and \wcircle{2} R8, resulting in a total of 110 APKs: 10 original non-obfuscated apps, 90 variants generated with \textsc{Obfuscapk}, and 10 variants generated with R8. We include R8 because it is the default Android optimization and obfuscation tool integrated into Android Studio and widely adopted in practice, as confirmed by previous studies~\cite{niroshan2025empirical}.

From the 21 transformations provided by \textsc{Obfuscapk}, we excluded plugins targeting resources, libraries, APK rebuilding, or manifest manipulation. We further excluded \textit{AdvancedReflection}, due to its overlap with \textit{Reflection}, and \textit{Nop}, since it only inserts semantically irrelevant no-operation instructions. After applying these exclusion criteria, nine \textsc{Obfuscapk} transformations remained: \textit{ArithmeticBranch}, \textit{CallIndirection}, \textit{ClassRename}, \textit{ConstStringEncryption}, \textit{FieldRename}, \textit{Goto}, \textit{MethodOverload}, \textit{MethodRename}, and \textit{Reflection}, together with R8 as a separate obfuscation configuration. Detailed descriptions of these transformations are available in the official \textsc{Obfuscapk} repository\footnote{\url{https://github.com/Mobile-IoT-Security-Lab/Obfuscapk}}.

We excluded obfuscators such as \textsc{DexGuard}, \textsc{DashO}, and \textsc{Allatori} due to their commercial nature. Among these tools, only \textsc{Allatori} provides a publicly accessible trial version. Nevertheless, the obfuscation produced by the trial version is immediately recognizable. In particular, it often introduces highly characteristic naming patterns such as repeated strings like \texttt{ALLATORI}, \texttt{m9046ALLATORIxDEMO}, or similarly distinctive method and class identifiers, which rarely appear in legitimate production builds.
Furthermore, the exclusion of \textsc{DashO} is unlikely to substantially affect the representativeness of our benchmark, as Niroshan et al.~\cite{niroshan2025empirical} found that only 1.01\% of the Google Play apps analyzed in their study were obfuscated using \textsc{DashO}, indicating limited adoption in practice.

\noindent
\textbf{\bcircle{2} The second dataset, $D_{W}$}, consists of 1000 randomly selected Android APKs collected from the Google Play Store via the AndroZoo repository in 2026. Unlike $D_{GT}$, which provides a controlled benchmark with known obfuscation transformations, $D_{W}$ is used to evaluate the capability of LLMs to detect obfuscation in large-scale real-world scenarios. 
We use a random sample instead of only popular apps to reduce selection bias and make the dataset more representative of the broader Google Play ecosystem. If we kept only popular apps, the results would be skewed toward apps with higher download counts, stronger developer resources, and often more mature security practices, which could inflate or distort obfuscation estimates compared with the real-world app population

\subsection{Metrics.}
\label{sec:metrics}
In this section, we define the evaluation metrics used to assess the effectiveness of LLMs in both binary obfuscation detection and obfuscation-technique classification.

\noindent
\textbf{Binary Obfuscation Detection Metrics.}
For our obfuscation detection analysis, we define the following primary outcomes:
\begin{itemize}
\item \textit{True Positives (TPs)}: The cases in which an application is obfuscated and the LLM correctly classifies it as obfuscated.
\item \textit{False Positives (FPs)}: The cases in which an application is not obfuscated (clean) but the LLM classifies it as obfuscated.
\item \textit{True Negatives (TNs)}: The cases in which an application is not obfuscated and the LLM correctly classifies it as non-obfuscated.
\item \textit{False Negatives (FNs)}: The cases in which an application is obfuscated but the LLM classifies it as non-obfuscated.
\end{itemize}

\noindent
Because the dataset is highly imbalanced, comprising 110 obfuscated samples and only 10 clean samples, we intentionally avoid relying on metrics such as \textit{Accuracy}, \textit{Precision}, and \textit{F1-score}. Under this imbalance, metrics dominated by the positive class may provide a misleading view of performance. For example, a trivial classifier labeling every app as obfuscated would still achieve approximately (91.6\%) \textit{Accuracy} and \textit{Precision}, despite being completely unable to recognize a clean app. For this reason, we focus on metrics that better capture the trade-off between detection capability and erroneous alarms. In particular, we consider \textit{Recall}:
\begin{equation}
Recall = \frac{TP}{TP + FN}
\end{equation}
which measures the ability of the model to correctly identify obfuscated applications and \textit{False Positive Rate}:
\begin{equation}
FPR = \frac{FP}{FP + TN}
\end{equation}
which quantifies the proportion of clean applications incorrectly classified as obfuscated. Our objective is to maximize Recall while simultaneously minimizing the FPR. Since these two goals are inherently competing, we additionally rely on \textit{Youden's J} statistic~\cite{youden1950index}, which combines both measures into a single coefficient:
\begin{equation}
Youden's\ J = Recall - FPR
\end{equation}
A higher $J$ value indicates a better balance between detection capability and false positive control, rewarding configurations that achieve high sensitivity while avoiding excessive erroneous alarms on clean 

\noindent
\textbf{Obfuscation Technique Classification Metrics.}
For the multi-class obfuscation technique classification task (see RQ2), we use \textit{Top-1 Accuracy} and \textit{Top-3 Accuracy} as our primary evaluation metrics. \textit{Top-1 Accuracy} measures the proportion of samples for which the highest-ranked prediction produced by the model matches the ground-truth obfuscation technique. \textit{Top-3 Accuracy} relaxes this requirement and considers a prediction correct if the ground-truth technique appears among the three highest-ranked predictions. Both metrics are computed as:
\begin{equation}
\text{Top-}k\text{ Accuracy} =
\frac{\#\;correct\;predictions}{\#\;total\;predictions}
\end{equation}
where a prediction is considered correct when the ground-truth obfuscation technique appears among the top-$k$ predictions returned by the model, with $k=1$ for Top-1 Accuracy and $k=3$ for Top-3 Accuracy. These metrics allow us to evaluate both the model's ability to identify the correct obfuscation technique as its most likely prediction and its ability to include the correct technique among its most plausible hypotheses.
\section{Evaluation.}
\label{sec:results}
In this section, we present the results of our empirical evaluation of LLMs for obfuscation detection and classification. We organize our evaluation around four research questions (RQs). The first three are investigated on the controlled benchmark $D_{GT}$, which provides ground-truth labels for both obfuscated and clean applications.

\begin{itemize}[leftmargin=*]
\item \textbf{RQ1:} Can LLMs reliably detect whether an Android app is obfuscated?
\item \textbf{RQ2:} Can LLMs identify the specific obfuscation technique applied to an app?
\item \textbf{RQ3} How do LLMs compare to existing obfuscation detection tools?
\end{itemize}

\noindent
The last RQ is evaluated on the real-world dataset $D_{W}$, where Ground-truth labels are not available a priori.
\begin{itemize}[leftmargin=*]
\item \textbf{RQ4} How do LLMs perform on Android apps sourced from Google Play?
\end{itemize}

\subsection{RQ1: Binary Obfuscation Detection.}
\label{sec:rq1}
To answer RQ1, and identify the most effective combination of LLM, prompt formulation, and decision threshold $T$, we evaluated all configurations on the controlled dataset $D_{GT}$ introduced in Section~\ref{sec:datasets}. In particular, we compared six LLMs across the three prompt variants ($prompt_{v1}$, $prompt_{v2}$, and $prompt_{v3}$) and nine threshold values ranging from 0.1 to 0.9. As discussed in Section~\ref{sec:metrics}, we rely on Youden’s J coefficient to evaluate the trade-off between Recall and False Positive Rate (FPR), allowing us to identify the configurations that achieve the best balance between detection capability and false alarm control. Table \ref{tab:RQ1} reports the performance of the evaluated LLMs across different thresholds using Youden’s J coefficient.

\begin{table*}[ht]
\centering
\scriptsize
\setlength{\tabcolsep}{4pt}

\caption{Youden's \(J\) coefficient for each model across nine decision thresholds (\(0.1\)--\(0.9\)) and three prompt versions (\(prompt_{v1}\), \(prompt_{v2}\), \(prompt_{v3}\)). The table reports the discriminative performance of each model at each threshold, where higher values indicate a better balance between sensitivity and specificity. Yellow highlights the best result in each row, cyan highlights the best result in each column, and green marks the intersection of both criteria. Values close to zero or below zero indicate weak or no class separation.}
\label{tab:RQ1}

\begin{tabular}{llccccccccc}
\toprule
& & \multicolumn{9}{c}{\textbf{Thresholds}} \\
\cmidrule(lr){3-11}

\textbf{Prompt}
& \textbf{Model}
& \textbf{0.1}
& \textbf{0.2}
& \textbf{0.3}
& \textbf{0.4}
& \textbf{0.5}
& \textbf{0.6}
& \textbf{0.7}
& \textbf{0.8}
& \textbf{0.9} \\
\midrule

\multirow{6}{*}{$prompt_{v1}$}
& gpt-4o-mini
&\cellcolor{yellow!25} 0.26
& 0.22
& 0.20
& 0.17
& 0.14
& 0.11
& 0.10
& 0.10
& 0.09 \\

& gpt-5-mini
&\cellcolor{yellow!25} 0.71
& 0.64
& 0.61
& 0.51
& 0.46
& 0.33
& 0.26
& 0.13
& 0.10 \\

& deepseek-r1\_32b
& 0.00
& 0.00
& 0.00
& 0.00
& 0.00
& 0.14
& 0.16
& \cellcolor{yellow!25}0.19
& 0.00 \\

& gemma3\_27b
& 0.00
& -0.02
& 0.12
& 0.21
&\cellcolor{yellow!25} 0.32
&\cellcolor{yellow!25} 0.32
& 0.24
& 0.15
& 0.06 \\

& qwen3\_30b
& 0.00
& -0.08
& -0.05
&\cellcolor{yellow!25} 0.01
& 0.00
& 0.00
& 0.00
& 0.00
& 0.00 \\

& gpt-oss\_20b
& \cellcolor{yellow!25}0.48
& 0.43
& 0.32
& 0.30
& 0.18
& 0.14
& 0.11
& 0.10
& 0.10 \\

\midrule

\multirow{6}{*}{$prompt_{v2}$}
& gpt-4o-mini
&\cellcolor{yellow!25} 0.44
& 0.30
& 0.21
& 0.19
& 0.17
& 0.14
& 0.12
& 0.10
& 0.10 \\

& gpt-5-mini
& \cellcolor{yellow!25} 0.73
& 0.65
& 0.63
& 0.56
& 0.48
& 0.39
& 0.30
& 0.16
& 0.11 \\

& deepseek-r1\_32b
& 0.00
& 0.00
& 0.00
& 0.00
& -0.01
& \cellcolor{yellow!25} 0.16
& 0.12
& 0.12
& 0.00 \\

& gemma3\_27b
& 0.00
& -0.01
& 0.16
& 0.20
& \cellcolor{yellow!25}0.36
& 0.23
& 0.22
& 0.12
& 0.05 \\

& qwen3\_30b
& -0.01
& \cellcolor{yellow!25}0.13
& -0.11
& 0.00
& 0.00
& 0.00
& 0.00
& 0.00
& 0.00 \\

& gpt-oss\_20b
& \cellcolor{yellow!25}0.48
& 0.44
& 0.41
& 0.37
& 0.25
& 0.18
& 0.13
& 0.11
& 0.11 \\

\midrule

\multirow{6}{*}{$prompt_{v3}$}
& gpt-4o-mini
& 0.66
&\cellcolor{yellow!25} 0.71
& 0.57
& 0.47
& 0.37
& 0.29
& 0.24
& 0.19
& 0.09 \\

& gpt-5-mini
&\cellcolor{green!25} 1.00
& 0.73
& 0.65
& \cellcolor{cyan!25}0.63
&  \cellcolor{cyan!25}0.56
& \cellcolor{cyan!25} 0.48
& 0.39
&\cellcolor{cyan!25} 0.30
& \cellcolor{cyan!25} 0.16 \\

& deepseek-r1\_32b
& 0.00
& 0.00
& 0.00
& 0.00
& 0.00
& -0.01
&\cellcolor{green!25} 0.46
& 0.12
& 0.00 \\

& gemma3\_27b
& 0.00
& -0.03
& 0.26
& 0.26
&\cellcolor{yellow!25} 0.44
& 0.30
& 0.24
& 0.12
& -0.01 \\

& qwen3\_30b
& 0.00
& -0.12
&\cellcolor{yellow!25} 0.04
& 0.01
& 0.00
& 0.00
& 0.00
& 0.00
& 0.00 \\

& gpt-oss\_20b
&\cellcolor{yellow!25} 0.77
& \cellcolor{cyan!25}0.74
& \cellcolor{cyan!25}0.69
& 0.61
& 0.52
& 0.39
& 0.27
& 0.17
& 0.11 \\

\bottomrule
\end{tabular}
\end{table*}

Overall, the results reveal a strong dependency on both the selected threshold and the prompt formulation. Across all configurations, low thresholds ($T=[0.1,0.3]$) consistently provide the best balance between detection capability and false positive control. As the threshold increases, Recall decreases significantly for most models, leading to lower overall J values. Among the evaluated models, \texttt{gpt-5-mini} consistently achieves the highest balanced performance across all three prompts, particularly at low thresholds ($T \leq 0.3$), peaking at $J = 1.00$ with $prompt_{v3}$ at $T = 0.1$. \texttt{gpt-oss\_20b} represents the strongest open-weight alternative, reaching $J = 0.77$ under the same conditions. \texttt{gpt-4o-mini} exhibits a stable but more moderate profile across all configurations. In contrast, \texttt{gemma3\_27b} tends to perform better only at mid-range thresholds ($T \approx 0.5$), while \texttt{deepseek-r1\_32b} and \texttt{qwen3\_30b} show consistently weak or unstable results across most settings.
Regarding prompt design, $prompt_{v3}$, which explicitly enumerates obfuscation techniques, yields the strongest and most consistent results across models, suggesting that detailed domain-specific context meaningfully guides the model's reasoning. $prompt_{v1}$ and $prompt_{v2}$ follow similar patterns but with uniformly lower peak values. With respect to the threshold $T$, lower values ($T \leq 0.2$) generally favor high Recall at the cost of increased false positives, while higher values reduce sensitivity without proportional gains in specificity. The optimal operating point for most competitive models lies in the range $T \in [0.1, 0.2]$.

\begin{figure}[ht]
\centering
\begin{subfigure}{\linewidth}
    \centering
    \includegraphics[width=\linewidth]{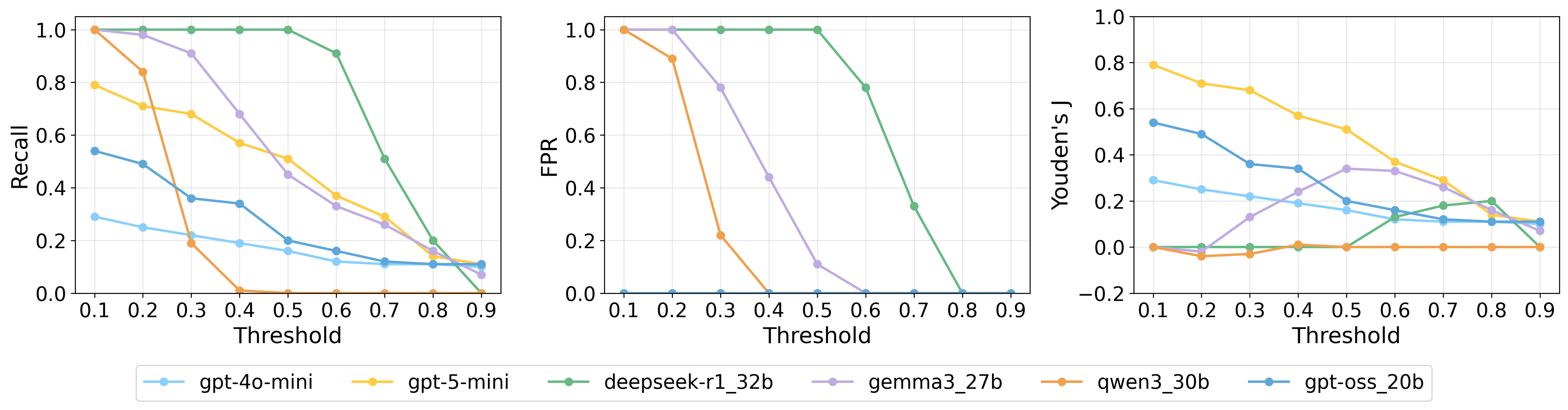}
    \caption{Performance metrics across thresholds using $prompt_{v1}$.}
    \label{fig:RQ1_ResultsV1}
\end{subfigure}
\vspace{0.5cm}
\begin{subfigure}{\linewidth}
    \centering
    \includegraphics[width=\linewidth]{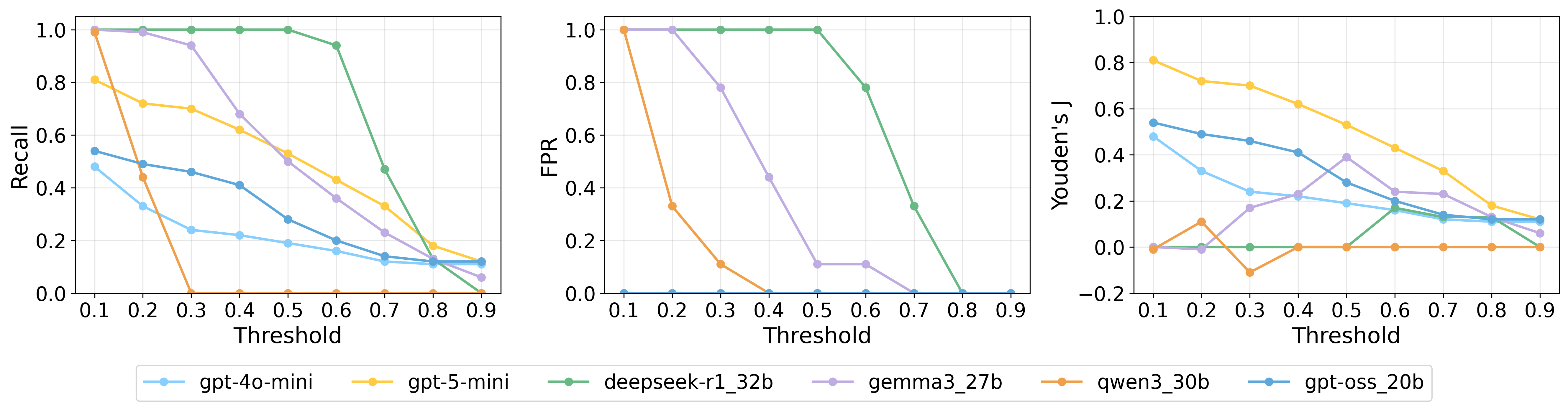}
    \caption{Performance metrics across thresholds using $prompt_{v2}$.}
    \label{fig:RQ1_ResultsV2}
\end{subfigure}
\vspace{0.5cm}
\begin{subfigure}{\linewidth}
    \centering
    \includegraphics[width=\linewidth]{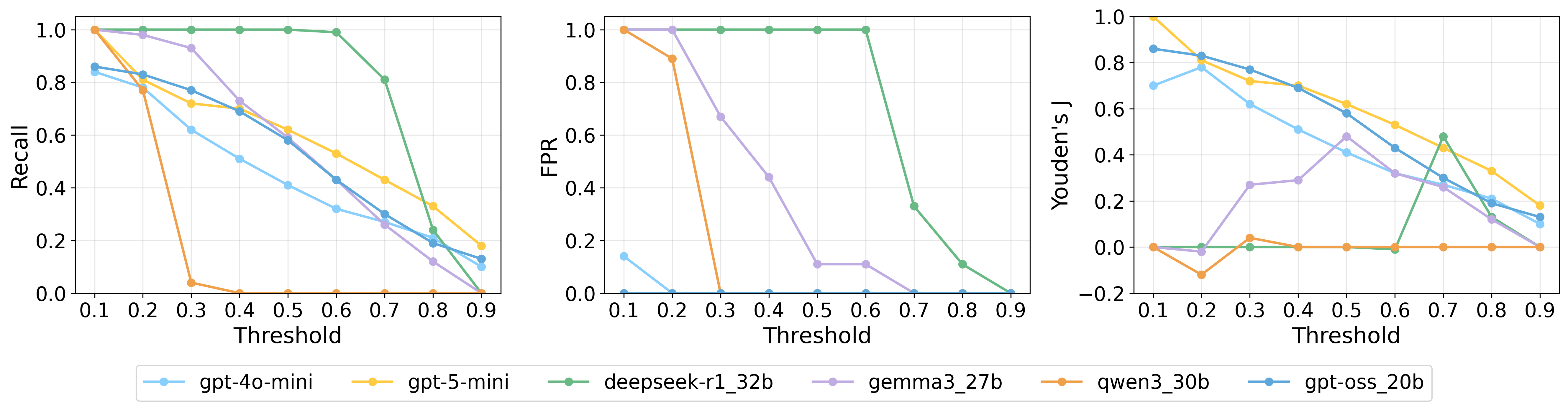}
    \caption{Performance metrics across thresholds using $prompt_{v3}$.}
    \label{fig:RQ1_ResultsV3}
\end{subfigure}
\caption{Recall, False Positive Rate (FPR), and Youden’s J coefficient obtained by the evaluated LLMs across different threshold values and prompt formulations.}
\label{fig:RQ1_Results_All}
\end{figure}

\noindent
\textbf{Relationship between Recall and FPR.}
Figure~\ref{fig:RQ1_Results_All} further decomposes performance into Recall and FPR, revealing an important nuance hidden by the aggregate J scores. The plots highlight that some models, such as \texttt{deepseek-r1\_32b}, achieve extremely high Recall values across a broad range of thresholds. However, this behavior is accompanied by an extremely high FPR, meaning that the model classifies nearly every application as obfuscated. While such behavior maximizes sensitivity, it makes the model impractical for realistic deployment because it fails to distinguish between clean and obfuscated apps. Conversely, \texttt{gpt-5-mini} and \texttt{gpt-oss\_20b} maintain high Recall while keeping FPR substantially lower, demonstrating a genuine ability to discriminate between obfuscated and clean applications. This distinction further underlines why Youden's J, rather than Recall alone, is the appropriate metric for this task. Overall, the results indicate that combining low thresholds with detailed prompt engineering provides the most reliable configuration for binary obfuscation detection.

\begin{figure}[ht]
\centering
\includegraphics[width=0.9\linewidth]{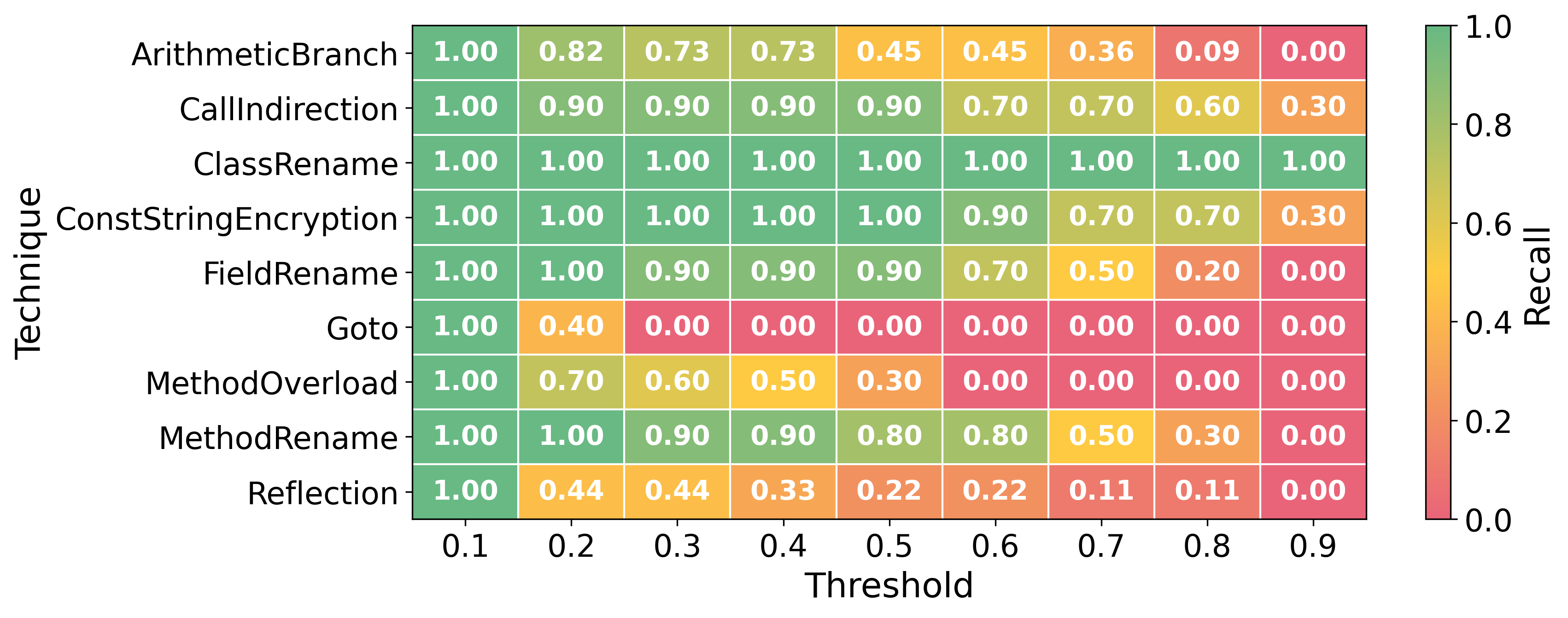}
\caption{Recall obtained by LLM for each obfuscation technique across different threshold values using \texttt{gpt-5-mini} and \texttt{promptv3}.}
\label{fig:RQ1_RecallByTechnique}
\end{figure}

\noindent
\textbf{Recall by Obfuscation Technique.}
While the aggregate Recall and FPR values provide a global view of the detection capability of the evaluated models, they do not capture how effectively the methodology handles specific obfuscation transformations. To better understand these differences, we further decompose the analysis by reporting the Recall obtained for each obfuscation technique individually.
Since evaluating all combinations of models and prompts would result in a large number of highly similar plots, we report only a representative configuration using \texttt{gpt-5-mini} with \texttt{promptv3}, which achieved the highest overall performance in the binary detection analysis. Nevertheless, the same qualitative trends are consistently observed across the other evaluated LLMs and prompt formulations, indicating that certain obfuscation techniques are systematically easier to detect than others, independently of the selected model. Figure~\ref{fig:RQ1_RecallByTechnique} presents a heatmap of the Recall achieved for each obfuscation technique across different threshold values using \texttt{gpt-5-mini} and \texttt{promptv3}.

The results reveal a clear divergence in detection difficulty across techniques. Identifier-based transformations such as \texttt{ClassRename}, \texttt{MethodRename}, and \texttt{FieldRename} achieve consistently high Recall across a broad range of thresholds, suggesting that their pervasive and distinctive naming patterns provide a strong signal for the model. In contrast, techniques such as \texttt{Goto}, \texttt{ArithmeticBranch}, and \texttt{Reflection} exhibit substantially lower Recall, particularly at higher thresholds. This behavior may be related to how these techniques are applied at the class level: since they alter only a limited subset of instructions within a class, the resulting obfuscation signal is weaker and harder for the LLM to detect confidently. These differences will be further discussed in Section~\ref{sec:discussion}. 
Please note that R8 is excluded from this plot as it simultaneously combines multiple transformations; however, its Recall is consistent with those observed for \texttt{MethodRename} and \texttt{ClassRename}, which represent its primary constituent techniques.

\begin{answer}
\textbf{Answer to RQ1:} LLMs can effectively detect obfuscation in Android apps when guided with sufficiently detailed prompts and appropriate decision thresholds. Among the evaluated configurations, $prompt_{v3}$ combined with low thresholds ($T=0.1$–$0.3$) achieves the best trade-off between Recall and FPR. Proprietary models, particularly \texttt{gpt-5-mini}, consistently provide the strongest and most stable performance, while some open-weight models exhibit high Recall at the cost of excessive false positives. Overall, the results demonstrate that LLM-based semantic reasoning represents a viable and effective approach for binary obfuscation detection in Android apps.
\end{answer}

\subsection{RQ2: Obfuscation Technique Classification.}
\label{sec:rq2}
To answer RQ2, we evaluate whether LLMs are capable of correctly identifying the specific obfuscation technique applied using the controlled dataset ($D_{GT}$). Classification performance is assessed using the Top-1 and Top-3 Accuracy metrics defined in Section~\ref{sec:metrics}. Table~\ref{tab:rq2} reports the classification performance achieved by the evaluated LLMs.

\begin{table*}[ht]
\centering
\caption{Per-technique obfuscation classification accuracy. T-1 (top-1) and T-3 (top-3) indicate whether the ground-truth technique appears among the top-1 or top-3 model predictions, respectively. The best-performing model for each technique under T-1 is highlighted in yellow, while the best under T-3 is highlighted in cyan.}
\label{tab:rq2}
\begin{adjustbox}{width=\linewidth}
\begin{tabular}{lcccccccccccc}
\toprule

& \multicolumn{2}{c}{\textbf{gpt-4o-mini}}
& \multicolumn{2}{c}{\textbf{gpt-5-mini}}
& \multicolumn{2}{c}{\textbf{deepseek-r1}}
& \multicolumn{2}{c}{\textbf{gemma3}}
& \multicolumn{2}{c}{\textbf{qwen3}}
& \multicolumn{2}{c}{\textbf{gpt-oss}} \\

\cmidrule(lr){2-3}
\cmidrule(lr){4-5}
\cmidrule(lr){6-7}
\cmidrule(lr){8-9}
\cmidrule(lr){10-11}
\cmidrule(lr){12-13}

\textbf{Technique}
& \textbf{T-1} & \textbf{T-3}
& \textbf{T-1} & \textbf{T-3}
& \textbf{T-1} & \textbf{T-3}
& \textbf{T-1} & \textbf{T-3}
& \textbf{T-1} & \textbf{T-3}
& \textbf{T-1} & \textbf{T-3} \\

\midrule

ArithmeticBranch
& 0.82 & \cellcolor{cyan!25}1.00
& \cellcolor{yellow!25}0.91 &\cellcolor{cyan!25} 1.00
& 0.55 & 0.91
& 0.36 & \cellcolor{cyan!25}1.00
&\cellcolor{yellow!25} 0.91 &\cellcolor{cyan!25} 1.00
&\cellcolor{yellow!25} 0.91 &\cellcolor{cyan!25} 1.00 \\

CallIndirection
& 0.10 & 0.80
& \cellcolor{yellow!25}0.80 &\cellcolor{cyan!25} 1.00
& 0.00 & 0.00
& 0.00 & 0.10
& 0.00 & 0.20
& 0.00 &\cellcolor{cyan!25} 1.00 \\

ClassRename
& \cellcolor{yellow!25}1.00 & \cellcolor{cyan!25}1.00
& \cellcolor{yellow!25}1.00 & \cellcolor{cyan!25}1.00
& \cellcolor{yellow!25}1.00 & \cellcolor{cyan!25}1.00
& \cellcolor{yellow!25}1.00 & \cellcolor{cyan!25}1.00
& \cellcolor{yellow!25}1.00 & \cellcolor{cyan!25}1.00
& \cellcolor{yellow!25}1.00 & \cellcolor{cyan!25}1.00 \\

ConstStringEncryption
& 0.90 & \cellcolor{cyan!25}1.00
&\cellcolor{yellow!25} 1.00 &  \cellcolor{cyan!25}1.00
& 0.90 &  \cellcolor{cyan!25}1.00
& 0.90 & \cellcolor{cyan!25} 1.00
& 0.90 &\cellcolor{cyan!25} 1.00
& \cellcolor{yellow!25}1.00 & \cellcolor{cyan!25}1.00 \\

FieldRename
& 0.80 & 0.90
&\cellcolor{yellow!25} 0.90 &  \cellcolor{cyan!25}1.00
& 0.50 & 0.80
& 0.20 & 0.90
& 0.80 & \cellcolor{cyan!25}1.00
& 0.80 &\cellcolor{cyan!25} 1.00 \\

Goto
& \cellcolor{yellow!25}1.00 &\cellcolor{cyan!25} 1.00
& \cellcolor{yellow!25}1.00 &\cellcolor{cyan!25} 1.00
& \cellcolor{yellow!25}1.00 & \cellcolor{cyan!25}1.00
& 0.70 & 1.00
& \cellcolor{yellow!25}1.00 &\cellcolor{cyan!25} 1.00
& \cellcolor{yellow!25}1.00 &\cellcolor{cyan!25} 1.00 \\

MethodOverload
& 0.70 & 1.00
& \cellcolor{yellow!25}0.80 & \cellcolor{cyan!25}1.00
& 0.40 & 0.70
& 0.10 & 0.50
& 0.70 &\cellcolor{cyan!25} 1.00
& 0.70 &\cellcolor{cyan!25} 1.00 \\

MethodRename
& 0.80 & 1.00
&0.80 & \cellcolor{cyan!25}1.00
& 0.60 &\cellcolor{cyan!25} 1.00
& \cellcolor{yellow!25} 0.90 &\cellcolor{cyan!25} 1.00
& 0.80 & \cellcolor{cyan!25}1.00
& 0.80 & \cellcolor{cyan!25}1.00 \\

Reflection
& 0.22 & \cellcolor{cyan!25}0.56
& \cellcolor{yellow!25} 0.33 &  0.44
& 0.22 & 0.44
& 0.00 & 0.11
& 0.22 & 0.33
& \cellcolor{yellow!25}0.33 & 0.44 \\

\midrule

\textbf{Overall}
& 0.71 & 0.92
& \cellcolor{yellow!25} 0.84 & \cellcolor{cyan!25} 0.94
& 0.58 & 0.77
& 0.47 & 0.74
& 0.71 & 0.84
& 0.73 & \cellcolor{cyan!25} 0.94 \\

\bottomrule
\end{tabular}
\end{adjustbox}
\end{table*}

The results indicate that LLMs are capable of recognizing specific obfuscation transformations with high accuracy, although substantial differences emerge across both models and techniques. Furthermore, the gap between top-1 and top-3 accuracy for several techniques indicates that, even when the top prediction is incorrect, the correct obfuscation strategy is often included among the model's most plausible hypotheses.

Among all evaluated models, \texttt{gpt-5-mini} achieves the strongest overall performance, obtaining an average top-1 accuracy of 0.84 and a top-3 accuracy of 0.94. The model demonstrates near-perfect classification capability for several transformations, including ClassRename, ConstStringEncryption, and Goto, while maintaining strong performance on MethodRename, MethodOverload, and FieldRename. In contrast, Reflection appears to be the most challenging technique to identify, reaching only 0.33 top-1 accuracy and 0.44 top-3 accuracy. This result is consistent with our findings from RQ1 (see Section~\ref{sec:rq1}), where Reflection also exhibited lower detection rates than identifier-based obfuscation techniques.

Interestingly, some models, such as \texttt{deepseek-r1}, despite achieving relatively weak performance in the binary detection task (RQ1), demonstrate a stronger ability to classify the specific obfuscation technique once it has been recognized. The model achieves perfect top-1 accuracy for ClassRename and Goto, and high scores for ConstStringEncryption.

Overall, the results suggest that LLMs can go beyond simple binary obfuscation detection and provide meaningful insights into the nature of the applied transformation. This capability is particularly valuable for security analysts, as understanding the specific obfuscation strategy can facilitate subsequent reverse-engineering and security assessment activities.

\begin{answer}
\textbf{Answer to RQ2:}
LLMs can effectively identify the specific obfuscation techniques applied to Android apps. \texttt{gpt-5-mini} achieves the highest classification accuracy (top-1 = 0.84, top-3 = 0.94). While identifier-based transformations are consistently the easiest to recognize, obfuscation patterns with a limited syntactic footprint, such as Reflection and CallIndirection, remain the most challenging techniques to classify reliably.
\end{answer}

\subsection{RQ3: Comparison with State-of-the-Art Tools.}
\label{sec:rq3}
To answer RQ3, we compare our LLM-based methodology against existing state-of-the-art Android obfuscation detection tools. The goal is to evaluate whether semantic reasoning provided by LLMs can improve detection performance over traditional approaches.

In Section \ref{sec:relatedwork}, we performed a literature review and identified existing approaches for obfuscation detection. Among the 14 papers identified, only 10 provided publicly available code or tools to facilitate the reproducibility of their results, while the remaining four papers did not release any artifacts and therefore cannot be used for comparison. Among the 10 remaining papers, EspyDroid \cite{gajrani2020espydroid+}, DINA \cite{alhanahnah2020dina}, and StaDART \cite{ahmad2020stadart} are primarily designed for malware detection, where obfuscation is considered only as an auxiliary feature. Since our study focuses on the detection and characterization of obfuscation techniques, these approaches fall outside the scope of our evaluation and were therefore excluded.
For the remaining seven papers, we attempted to reproduce the corresponding approaches using the source code released by the authors. However, the repositories associated with Wang et al. \cite{wang2017changed}, AndrODet \cite{androdetRepo}, and Conti et al. \cite{ContiRepo} were non-functional at the time of evaluation. We contacted the respective authors to obtain support and clarification, but either received no response or were unable to resolve the execution issues. As a result, despite our best efforts, these approaches could not be reliably executed and were therefore excluded from the comparison. 
This limitation is outside our control and reflects broader reproducibility challenges that continue to affect a portion of the Android security research literature. Consequently, our comparative evaluation focuses on the four reproducible and readily executable tools that provide dedicated support for obfuscation detection: \wcircle{1} \textbf{SEBASTiAN}, \wcircle{2} \textbf{Trueseeing}, \wcircle{3} \textbf{APKHunt}, and \wcircle{4} \textbf{Niroshan et al.}. Regarding Niroshan et al., we contacted the authors directly to obtain assistance with running their obfuscation detection tool. Thanks to their kind collaboration and the technical guidance they provided, we were able to execute the tool reproducibly and include it in our empirical comparison.
Table \ref{tab:rq3} summarizes the comparative performance between our proposed methodology and tools available at the state-of-the-art, evaluated on the $D_{GT}$ dataset. 

\begin{table*}[ht]
\centering
\caption{Comparison between our LLM-based methodology and existing SAST-based obfuscation detection tools on $D_{GT}$. (*) denotes the best-performing LLM configuration identified in RQ1, corresponding to \textit{gpt-5-mini} with $prompt_{v3}$ and threshold $T=0.1$. Cells with a green background highlight the best result achieved in each column.}
\label{tab:rq3}

\vspace{1em}

\begin{adjustbox}{width=0.7\linewidth}
\begin{tabular}{lcccc}
\toprule
\textbf{Tool}
& \textbf{Recall}
& \textbf{FPR}
& \textbf{Youden's $J$} & \textbf{AVG Time} \\
\midrule
LLM-driven$^{*}$
&\cellcolor{green!25} 1.00
& \cellcolor{green!25}0.00
& \cellcolor{green!25}1.00 & 282.17 s \\
APKHunt
& 0.00
&\cellcolor{green!25}0.00
& 0.00  & 338.63 s\\
SEBASTiAN
& 0.00
& \cellcolor{green!25}0.00
& 0.00 & 137.89 s\\
Trueseeing
& 0.04
& 0.82
& -0.77 & \cellcolor{green!25}18.32 s \\
Niroshan et al.
& 0.78
& 0.9
& -0.11 &  126.40 \\
\bottomrule
\end{tabular}
\end{adjustbox}
\end{table*}

The results indicate that LLMs substantially outperform the evaluated tools (even when considering configurations other than the best-performing one identified in RQ1).
In particular, APKHunt and SEBASTiAN did not identify any of the obfuscated apps in our benchmark, resulting in a Recall of 0.00. While both tools include checks that can surface obfuscation-related indicators, the obfuscation techniques considered in our study were not effectively captured by their implemented detection rules. These results suggest that the current detection capabilities of these tools may be limited when confronted with commonly used obfuscation strategies.
Trueseeing exhibits a different behavior. While it is able to correctly identify a small fraction of obfuscated apps (Recall = 0.04), it also produces a very large number of false alarms, resulting in a False Positive Rate (FPR) of 0.82 and a negative Youden's $J$ coefficient. This indicates that the tool struggles to reliably distinguish obfuscated apps from non-obfuscated ones in our dataset. 
Niroshan et al., instead, achieve a relatively good Recall ($0.78$) in identifying obfuscated apps; however, this comes at the cost of a very high False Positive Rate ($0.90$). Interestingly, these results are somewhat at odds with those reported in their original paper, especially considering that their approach is the only evaluated approach specifically designed for Android obfuscation detection. We further discuss potential reasons for this discrepancy in Section~\ref{sec:discussion}.

Overall, these findings suggest that semantic reasoning provided by LLMs may offer a significant advantage over traditional rule-based and heuristic-driven approaches for Android obfuscation analysis. Unlike conventional SAST tools, which depend on predefined signatures or handcrafted detection rules, LLMs appear capable of analyzing code semantics more effectively, resulting in substantially stronger detection performance.

\noindent
\textbf{Time analysis.}
Table~\ref{tab:rq3} also reports the average execution time required by each approach. Overall, the LLM-driven pipeline introduces some computational overhead compared to traditional SAST tools (while still outperforming APKHunt in terms of execution time), but it significantly improves detection performance. In contrast, a tool like Trueseeing, while considerably faster (less than 20 s per app), exhibits a negative Youden’s J score, indicating poor discriminative capability. Furthermore, it is important to note that our study is empirical in nature, and the proposed pipeline is not intended as a production-ready tool. Additional optimizations and engineering refinements could further reduce execution time, as discussed in Section~\ref{sec:discussion}.

\begin{answer}
\textbf{Answer to RQ3:}
LLMs consistently outperformed the evaluated SAST tools on our benchmark. Existing tools showed limited effectiveness in detecting several common Android obfuscation techniques, whereas LLMs achieved high detection performance without relying on handcrafted rules.
\end{answer}

\subsection{RQ4: Real-World Obfuscation Detection.}

To answer RQ4, we evaluate the proposed methodology on the real-world dataset $D_W$ (see Section~\ref{sec:results}), which consists of 1000 real-world apps collected from Google Play. The objective of this RQ is to assess how the best-performing LLM configuration identified in RQ1 behaves in a realistic deployment scenario. Based on the results of RQ1, we select \textit{gpt-5-mini} combined with \textit{promptv3}. Table~\ref{tab:rq4} reports the distribution of apps classified as obfuscated and non-obfuscated under the different threshold values tested.

\begin{table*}[t]
\centering
\caption{Distribution of apps classified as obfuscated and non-obfuscated in the real-world dataset ($D_W$) across different decision thresholds. Results are obtained using the best-performing configuration identified in RQ1 (\texttt{gpt-5-mini} with \texttt{promptv3}).}
\label{tab:rq4}

\vspace{1em}

\begin{adjustbox}{width=0.8\linewidth}
\begin{tabular}{ccc}
\toprule
\textbf{Threshold ($T$)} & \textbf{Obfuscated Apps (\%)} & \textbf{Not Obfuscated Apps (\%)} \\
\midrule
0.1 & 72\% & 28\% \\
0.2 & 63\% & 37\% \\
0.3 & 58\% & 42\% \\
0.4 & 54\% & 46\% \\
0.5 & 51\% & 49\% \\
0.6 & 46\% & 54\% \\
0.7 & 40\% & 60\% \\
0.8 & 37\% & 63\% \\
0.9 & 28\% & 72\% \\
\bottomrule
\end{tabular}
\end{adjustbox}
\end{table*}

Using the optimal threshold identified in RQ1 (\(T = 0.1\)), 72\% of the apps are classified as obfuscated, while the remaining 28\% are classified as non-obfuscated. This result is consistent with the finding of Niroshan et al.\cite{niroshan2025empirical}, who reported 66\% in 2023, and may indicate a plausible increase over time, since our analysis focuses on newer apps, i.e., from 2026. However, unlike $D_{GT}$, the true obfuscation status of apps in $D_W$ is not known a priori. Consequently, the classification results alone are insufficient to evaluate the LLMs. To evaluate the correctness of the generated labels, we therefore perform a manual inspection of the apps.

\noindent
\textbf{Manual Inspection.}
The objective of the manual inspection is to establish a reliable ground truth by determining whether each app is obfuscated through manual reverse engineering and code inspection.
Given the large number of apps, performing this analysis on the entire $D_W$ dataset is extremely challenging. We therefore manually inspected a statistically significant random sample, which is commonly employed in empirical software engineering studies when manual inspection of the entire population is impractical~\cite{li2026automatically, guglielmi2024help, alecci2025damflow, zhang2025killing, wang2024makes,  alecci2026taskflow, alecci2024improving, alecci2026geotwins}. Using a confidence level of 95\% and a margin of error of 10\%, we obtained a sample size of 88 apps. To ensure a balanced evaluation, half of the sample was randomly selected from the apps classified as obfuscated by the LLM, while the remaining half was selected from those classified as non-obfuscated.

Two authors independently analyzed each APK without access to the LLM's predictions, thereby reducing the risk of automation bias. For each app, the annotators decompiled the app and performed static inspection of the corresponding Smali code, searching for evidence of obfuscation transformations. Based on this analysis, each app was labeled as either \textit{obfuscated} or \textit{non-obfuscated}. After the independent annotation phase, the authors compared their assessments. In cases of disagreement, the corresponding applications were jointly reviewed and discussed until a consensus was reached, resulting in a final manually validated label for each inspected app.  Table~\ref{tab} reports the performance of the best-performing configuration identified in RQ1 (\texttt{gpt-5-mini} with \texttt{promptv3}) on the manually inspected subset of $D_W$, where the labels produced by the LLM were compared against the manually validated ground truth established during the inspection phase.

\begin{table*}[t]
\centering
\caption{Performance of the best-performing LLM configuration (\texttt{gpt-5-mini} with \texttt{promptv3}) on the manually inspected subset of $D_W$.}
\label{tab:rq4_metrics}

\vspace{1em}

\begin{adjustbox}{width=0.5\linewidth}
\begin{tabular}{lccc}
\toprule
 & \textbf{Precision} & \textbf{Recall} & \textbf{F1-score} \\
\midrule
\textbf{Score} & 0.82 & 0.96 & 0.88 \\
\bottomrule
\end{tabular}
\end{adjustbox}
\end{table*}

Overall, the results indicate that the proposed methodology maintains a high level of effectiveness when applied to real-world Android apps. The approach achieves a Precision of 0.82, a Recall of 0.96, and an F1-score of 0.88. These results suggest that the LLM is able to correctly identify the vast majority of obfuscated apps while keeping the number of false positives relatively limited. In particular, the very high Recall demonstrates that only a small fraction of obfuscated apps remain undetected.

The strong performance observed on real-world applications is consistent with the findings of RQ1. As discussed previously, LLMs are particularly effective at identifying renaming-based obfuscation techniques, such as class, method, and field renaming. This observation is especially relevant because recent studies have shown that identifier renaming represents one of the most prevalent obfuscation strategies adopted in Android applications, largely due to the widespread use of tools such as R8~\cite{niroshan2025empirical}. Consequently, the prevalence of these transformations in real-world apps likely contributes to the strong effectiveness achieved by the LLM in this evaluation. These findings suggest that the capability of LLMs to reason about obfuscation patterns generalizes beyond controlled benchmarks and remains effective when analyzing apps collected from the Google Play ecosystem.

\begin{answer}
\textbf{Answer to RQ4:}
The best-performing LLM configuration (\texttt{gpt-5-mini} with \texttt{promptv3}) generalizes well to real-world Android appss, achieving an F1-score of 0.88. These results indicate that LLMs can effectively detect obfuscated apps in realistic settings, particularly because the most common obfuscation techniques adopted in practice are renaming-based transformations, which LLMs identify with high accuracy.
\end{answer}

\section{Discussion}
\label{sec:discussion}
In this section, we discuss the main implications of our findings and their relevance for future research on Android obfuscation analysis.

\noindent
\textbf{Empirical nature of our study.}
Rather than proposing a new obfuscation detection tool, this paper investigates whether off-the-shelf LLMs can detect and classify Android obfuscation techniques without dedicated training or handcrafted rules. Consequently, the proposed pipeline should not be viewed as a production-ready tool for obfuscation detection or classification, but rather as a research prototype designed to explore the capabilities and limitations of current LLMs in this domain. The primary contribution of this work is therefore not the pipeline itself, but the insights it provides into the extent to which contemporary LLMs can support Android obfuscation analysis, as well as the challenges that remain.

\noindent
\textbf{LLMs are particularly effective on renaming-based obfuscation.}
Our findings suggest that obfuscation techniques are not equally easy to detect. Across the evaluated models, identifier-based transformations such as class, method, and field renaming consistently achieved the highest detection and classification rates. A likely explanation is that these techniques introduce strong and pervasive patterns that remain visible even after compilation to Smali code. 
This result is particularly relevant because identifier renaming is one of the most widely adopted obfuscation strategies in Android apps, and R8 (i.e., the default optimization and obfuscation tool integrated into Android Studio) relies heavily on renaming and shrinking transformations. Recent large-scale studies further confirm that R8 and related renaming-based obfuscation techniques are among the most prevalent in the Android ecosystem~\cite{niroshan2025empirical}.
Consequently, although LLMs are not equally effective across all obfuscation families, their strong performance on renaming-based transformations suggests that they can already provide meaningful benefits for a large fraction of obfuscated apps encountered in practice. By contrast, more sophisticated transformations may require deeper semantic reasoning or additional contextual signals to be identified with comparable accuracy.

\noindent
\textbf{Implications for obfuscation analysis.}
Our results suggest that LLMs reason primarily about the characteristics of the applied transformation rather than the specific obfuscation tool that produced it. This is particularly encouraging because many Android obfuscators, both open-source and commercial, rely on a common set of transformations, including identifier renaming, string encryption, and control-flow manipulation. For instance, commercial tools such as Allatori~\cite{allatori} employ renaming-based techniques that are conceptually similar to those generated by R8.
This suggests that the capabilities observed in our study may extend beyond the specific tools considered in the evaluation and apply to a broader range of Android obfuscation scenarios. More generally, our findings indicate that LLMs can reason about obfuscation patterns at the transformation level, which may make them less dependent on tool-specific signatures than traditional rule-based approaches.

\noindent
\textbf{Detection vs Classification.}
Finally, our results suggest that binary obfuscation detection and obfuscation classification are partially independent tasks. Some models, such as \texttt{deepseek-r1}, achieve relatively weak performance in binary detection due to a high false positive rate, yet still demonstrate a reasonable ability to identify the specific obfuscation technique when analyzing obfuscated code. This indicates that models may struggle to determine whether obfuscation is absent while still being able to characterize its nature once obfuscated code is analyzed.

\noindent
\textbf{Comparison with existing approaches.}
Niroshan et al.\ report up to 97\% Accuracy for their obfuscation detector, using Accuracy as the primary evaluation metric. However, as discussed in our metrics section \ref{sec:metrics}, such metrics become misleading under strong class imbalance: on our \(D_(GT)\) benchmark, a trivial classifier that always predicts an app as obfuscated would already reach 91.6\% Accuracy while never correctly identifying a clean app. For this reason, we focus on Recall, False Positive Rate, and Youden’s J, which more reliably capture the trade-off between detection capability and erroneous alarms. In addition, part of the dataset used by Niroshan et al.\ is obfuscated with the demo version of the Allatori obfuscator, whose characteristic identifier patterns (e.g., strings explicitly containing the tool name) are rarely observed in production builds and make obfuscation trivially detectable via simple pattern matching. By excluding the Allatori demo and relying instead on widely adopted tools such as R8 and Obfuscapk, our benchmark avoids such tool-specific artifacts and provides a more realistic assessment of obfuscation detection in real-world Android applications.
\section{Limitations}
\label{sec:limitations}
In this section, we discuss the main limitations affecting our study.

\noindent
\textbf{LLM non-determinism and reproducibility}
LLMs are inherently non-deterministic and may occasionally generate inconsistent or hallucinated outputs. These imperfections stem from the inherent limitations of current models rather than from our experimental design. To mitigate this threat, we carefully designed our prompts, ran each experiment 5 times with the temperature set to 0, and used the majority prediction as the final label. This procedure reduces output variability and improves the robustness of the reported results. Nevertheless, hallucinations and prediction inconsistencies cannot be completely eliminated and may still affect the full reproducibility of the analysis.

\noindent
\textbf{LLM selection constraints.} 
The study evaluates a representative set of contemporary open-weight and proprietary models based on our financial and hardware capabilities. However, the LLM ecosystem evolves rapidly, and new models may exhibit different capabilities. More advanced models, including proprietary systems (e.g., GPT-5.4 or Claude Opus 4.8), may yield improved performance, but their use introduces significantly higher operational costs and may limit reproducibility. As a result, our findings should not be interpreted as a comprehensive comparison of all existing LLMs, but rather as an empirical assessment of a diverse and widely adopted subset of current systems.

\noindent
\textbf{Benchmark Size.}
The controlled benchmark ($D_{GT}$) was intentionally designed to provide ground-truth labels, enabling the systematic evaluation of different LLM configurations. Although the benchmark comprises only 10 apps, its construction and evaluation required generating multiple obfuscated variants for each app and assessing them across several LLMs, prompt formulations, and repeated executions to mitigate model non-determinism. Consequently, even a relatively small benchmark produced a substantial number of LLM inferences, resulting in significant computational and financial costs. In addition, the development of the experimental pipeline, prompt engineering process, and preliminary experimentation required further LLM executions. As a result, expanding the benchmark to a substantially larger scale is extremely challenging. To mitigate this threat, we complemented the controlled benchmark with a large-scale real-world dataset collected from Google Play, allowing us to validate our findings beyond the controlled setting and assess the generalizability of the observed trends.

\noindent
\textbf{Bytecode-level obfuscation scope.}
Our analysis is performed on Smali code obtained from decompiled APKs, and therefore, our detector is fundamentally tailored to bytecode-level obfuscation. Accordingly, the reported results should be interpreted as detecting whether the Dalvik bytecode extracted from an app is obfuscated, rather than whether the entire app is obfuscated in a broader sense. This distinction is important because obfuscation may be applied to different parts of the app pipeline: for example, JavaScript or other higher-level components may be obfuscated before compilation, while the resulting Smali code may exhibit only the bytecode-level effects of those transformations. Future work could extend the proposed methodology beyond Smali code by incorporating the analysis of other app artifacts, such as JavaScript or resources, to provide a more comprehensive assessment of obfuscation across the entire Android app ecosystem.
\section{Conclusions}
\label{sec:conclusion}
This paper presents the first empirical study investigating the ability of LLMs to detect obfuscation in Android apps through semantic reasoning over Smali code. Our results show that LLMs can effectively detect obfuscated apps and identify specific obfuscation techniques without relying on handcrafted features, dedicated training, or predefined rules. Proprietary models generally achieve the strongest performance, outperforming traditional SAST-based approaches. Overall, our findings suggest that LLMs represent a promising direction for Android obfuscation analysis.
As future work, we plan to investigate the prevalence of obfuscation in Android malware and identify the most commonly used obfuscation techniques. We also plan to explore the use of LLMs for Android code deobfuscation, including their ability to recover meaningful identifiers, reconstruct higher-level program semantics, and support reverse-engineering activities.

\bibliography{references}

\appendix

\section{LLM Prompts}
\label{app:prompts}

\noindent
\textbf{Obfuscation Detection - $prompt_{v1}$:}

\begin{prompt}
\textbf{Prompt:}

You are a\textbf{ senior Android reverse engineer} specializing in smali and obfuscation analysis.

\textbf{Task:}    Determine whether the given smali class is obfuscated.

\textbf{Output rules:}

    - Respond with exactly one word: True or False

    - No explanations, no punctuation, no extra text!

\textbf{    Class Name:
}    \{className\}

\textbf{    Class Smali Code:
}    \{classContent\}

Now provide the Final Output [\textbf{True/False} only]!
\end{prompt}

\noindent
\textbf{Obfuscation Detection - $prompt_{v2}$:}

\begin{prompt}
\textbf{Prompt:} 

You are a \textbf{ senior Android reverse engineer} specializing in smali and obfuscation analysis.

\textbf{Task:}
    Determine whether the given smali class is obfuscated.

\textbf{Definition:}   

\textit{Obfuscation is any transformation that makes code harder for a human to understand, analyze, or reverse engineer while preserving its functionality. This applies to any part of the code: not only the method bodies, but also identifiers. If the semantics of the code are intentionally hidden or made impossible to infer, whether through altered logic or through loss of meaningful naming, the code can be considered obfuscated.}

Your task is to determine whether the given smali class shows signs of obfuscation based on this definition, without judging the degree of obfuscation.

\textbf{Output rules:}

    - Respond with exactly one word: True or False
    
    - No explanations, no punctuation, no extra text!

\textbf{Class Name:}    \{className\}

\textbf{Class Smali Code:} \{classContent\}

Now provide the Final Output [\textbf{True/False} only]!

\end{prompt}

\noindent
\textbf{Obfuscation Detection - $prompt_{v3}$:}

\begin{prompt}
\textbf{Prompt:}

You are a \textbf{senior Android reverse engineer} specializing in smali and obfuscation analysis.

\textbf{Task:}
    Determine whether the given smali class is obfuscated.

\textbf{Definition:}

\textit{Obfuscation is any transformation that makes code harder for a human to understand,analyze, or reverse engineer while preserving its functionality. This applies to any part of the code: not only the method bodies, but also identifiers. If the semantics of the code are intentionally hidden or made impossible to infer, whether through altered logic or through loss of meaningful naming, the code can be considered obfuscated. Naming obfuscation alone is sufficient to classify a class as obfuscated, even if the method bodies appear structurally simple or clean.}

The following techniques are representative examples of common obfuscation patterns, but this list is NOT exhaustive. Obfuscation may also involve custom techniques or combinations of multiple techniques.

The following techniques are representative examples of common obfuscation patterns, but this list is NOT exhaustive:

    - \textbf{ArithmeticBranch}: inserting arithmetic-based branches that are never taken

    - \textbf{CallIndirection}: wrapping method calls inside new methods

    - \textbf{ClassRename}: class or package names replaced with short, non-semantic identifiers (e.g. single letters, two-letter sequences, or meaningless alphanumeric tokens) such that the original purpose cannot be inferred from the name

    - \textbf{ConstStringEncryption}: encrypting constant strings

    - \textbf{FieldRename}: field names replaced with short or non-semantic identifiers

    - \textbf{Goto}: inserting redundant goto instructions altering control flow

    - \textbf{MethodOverload}: adding overloaded methods with extra arguments and junk bodies

    - \textbf{MethodRename}: method names replaced with short or non-semantic identifiers (e.g. single letters 
    a, b, c, d) such that the original purpose cannot be inferred from the name

    - \textbf{Reflection}: invoking methods via reflection instead of direct calls

Your task is to determine whether the given smali class shows signs of obfuscation, including but not limited to the techniques listed above.  Only classify as False if both the identifiers AND the method bodies show no signs of intentional obfuscation.

\textbf{Output rules:}

    - Respond with exactly one word: True or False
    
    - No explanations, no punctuation, no extra text

\textbf{Class Name:} \{className\}

\textbf{Class Smali Code:} \{classContent\}

Now provide the Final Output [\textbf{True/False} only]
\end{prompt}

\noindent
\textbf{Obfuscation Classification - $prompt_{classification}$}
\begin{prompt}

\textbf{Prompt:}    
You are a \textbf{senior Android reverse engineer} specializing in smali and obfuscation analysis.

\textbf{Task:}
Identify the PRIMARY obfuscation technique used in the class.

\textbf{Label set:} \{expectedLabels\}

\textbf{Label definitions:}

- \textbf{ArithmeticBranch}: inserting arithmetic-based branches that are never taken

- \textbf{CallIndirection}: wrapping method calls inside new methods

- \textbf{ClassRename}: renaming classes and packages

- \textbf{ConstStringEncryption}: encrypting constant strings

- \textbf{FieldRename}: renaming fields

- \textbf{Goto}: inserting redundant goto instructions altering control flow

- \textbf{MethodOverload}: adding overloaded methods with extra arguments and junk bodies

- \textbf{MethodRename}: renaming methods

- \textbf{Reflection}: invoking methods via reflection instead of direct calls

\textbf{Instructions:}

- Select the single most dominant technique

- If multiple techniques exist, choose the most evident or impactful one

- Ensure the label exactly matches one from the list

\textbf{Output rules:}

- Output exactly one label from the provided list

- No explanations, formatting, or additional text!

\textbf{Class Name:} \{className\}

\textbf{Class Smali Code:} \{classContent\}

Now provide the Final Output [one\textbf{ label }only]!
\end{prompt}

\end{document}